\documentstyle[11pt,epsfig]{article}
\begin{document}

\title{{\bf  Semiclassical corrections to the Einstein equation and Induced Matter Theory }}
\author{  P. Moyassari $^1$ \hspace{0.5 cm}and\hspace{0.5 cm} S. Jalalzadeh $^{2,3}$\thanks{%
email: s-jalalzadeh@sbu.ac.ir}\\
$^1${\small Department of physics, Tafresh University, Tafresh,
Iran.}\\
 $^2${\small Department of Physics, Shahid Beheshti University, Evin, Tehran
19839, Iran.}\\$^3${\small Research Institute for Astronomy and
Astrophysics of Maragha – Maragha, IRAN, P.O.Box: 55134-441.}}
\maketitle
\begin{abstract}

The induced Einstein equation on a perturbed brane in the Induced
Matter Theory is re-analyzed. We indicate that in a conformally
flat background, the local quantum corrections to the Einstein
equation can be obtained via the IMT. Using the FRW metric as the
4D gravitational model, we show that the classical fluctuations of
the brane may be related to the quantum corrections to the
classical Einstein equation. In other words, the induced Einstein
equation on the perturbed brane correspond with the semiclassical
Einstein equation.

PACS number: 04.50.+h, 04.62.+v, 03.65.Sq.
\end{abstract}

\vspace{2cm}
\section{Introduction}
One of the most fundamental problems in present-day physics concerns
a quantum theory of gravitation. In this context, the concept of
semiclassicality is particularly relevant in order to make it a
physical theory. Applying the semiclassical theory, as the first
order quantum corrections to the classical theory, one hopes to get
an insight into some of the structures of the full, elusive, theory.
In fact, the semiclassical approach provides the framework for some
realistic scenarios which may explain some of the features of the
present universe \cite{1a}. It would even allow us to say something
on quantum gravity without having to build the full consistent
theory. This is particularly interesting since experiments like
GLAST, AUGER and others, should be able to measure effects due to a
quantum gravity regime \cite{ba1}. In the semiclassical theory of
gravity a classical metric is coupled to the expectation value of
the stress tensor
\begin{equation}\label{0-1}
    G_{\mu\nu}=-8\pi G<T_{\mu\nu}>.
\end{equation}
It was found that this theory gives reliable results when the
fluctuations in the stress tensor are not too large \cite{2a}. The
semiclassical limit of the quantum gravity has to match the
established classical theory for consistency. The form of the
semiclassical corrections to the Einstein equation is known for many
important cases in the $4D$ spacetime \cite{3a}. In this paper we
will study the relation between the quantum corrections to the
Einstein equations and Induced Matter Theory (IMT) of Wesson. The
physics of gravitational interactions in higher dimensional
spacetime has received considerable interest in recent years. The
possibility that the fundamental Planck mass within a higher
dimensional setting may be as low as the electroweak scale
\cite{4a,5a,6a} has stimulated extensive model building and numerous
investigations aiming at signatures of extra dimensions. Central to
these scenarios is that gravity lives in higher dimensions, while
standard model particles are often confined to the four dimensional
brane. Higher dimensional extensions to general relativity were
originally started with works of Kaluza and Klein (KK) with the
addition of one extra dimension and subsequently generalized to more
extra dimensions by various authors. In KK theory, the components of
the $5D$ metric tensor is independent of the extra dimension. In
contrast to this model, Wesson suggested the IMT  \cite{9a} which
differs from KK theory by the fact that it has a noncompact
 fifth dimension and that the 5D bulk space is devoid of matter. In
 this theory the effective $4D$ matter is a consequence of the geometry
 of the $5D$ bulk space which is Ricci-flat while the $4D$ hypersurface is curved by the
 $4D$  induced matter\cite{10a}. One of the interesting futures of IMT is that it contains some quantum mechanical effects.
 For example it has been shown that particles and waves are merely different representations of the same underlying
  geometry and may be the same thing viewed in different ways
\cite{11a}. Also, using $5D$ IMT, a form of Heisenberg's relation
that applies to real and virtual particles has been derived by
Wesson \cite{12a}. We will show that the geometrical fluctuations
obtained via IMT may be related to the quantum mechanical effects.
In this paper we focus attention on IMT and proceed to derive the
semiclassical corrections to the Einstein equation using this
theory. In doing so, we will briefly review the geometrical
definitions and derive the induced Einstein equation on the
perturbed brane through contracting the Gauss-Codazzi equations. In
section three, we will consider a FRW universe as a non-perturbed
brane embedded in a $5D$ flat spacetime and study the Einstein
equation on the perturbed brane. We will show that these equations
correspond to the semiclassical Einstein equation. It means that the
classical fluctuations of the perturbed brane can be interpreted as
the quantum fluctuations of the matter field.

\section{Geometrical setup}

Consider the background manifold $\overline{V}_{4}$ isometrically
embedded in ${V}_{5}$ by a map ${\cal
Y}:\overline{V}_{4}\rightarrow V_{5}$ such that
\begin{equation}
{\cal G}_{AB}{\cal Y}_{\,\,\,,\mu }^{A}{\cal Y}_{\,\,\,,\nu }^{B}=\bar{g}%
_{\mu \nu },\hspace{0.5cm}{\cal G}_{AB}{\cal Y}_{\,\,\ ,\mu }^{A}{\cal N}%
^{B}=0,\hspace{0.5cm}{\cal G}_{AB}{\cal N}^{A}{\cal N}%
^{B}=1 \label{1-1}
\end{equation}%
where ${\cal G}_{AB}$ $(\bar{g}_{\mu \nu })$ is the metric of the
bulk (brane) space $V_{5}(\overline{V}_{4})$ in an arbitrary coordinate, $\{{\cal %
Y}^{A}\}$ $(\{x^{\mu }\})$ are the bases of the bulk (brane) and ${\cal N}%
^{A}$ is a normal unit vector orthogonal to the brane. Perturbation
of $\overline{V}_{4}$ in a sufficiently small neighborhood of the
brane along an arbitrary transverse direction $\xi$ is given by
\begin{equation}
{\cal Z}^{A}(x^{\mu },\zeta)={\cal Y}^{A}+({\cal L}_{\xi }{\cal
Y})^{A}, \label{1-2}
\end{equation}%
where ${\cal L}$ represents the Lie derivative and $\zeta$ is a
small parameter along ${\cal N}^{A}$ parameterizing the extra
noncompact dimension. By choosing $\xi$ orthogonal to the brane we
ensure gauge independency \cite{13a} and have perturbations of the
embedding along a single orthogonal extra direction $\bar{{\cal
N}}$, giving the local coordinates of the perturbed brane as
\begin{equation}
{\cal Z}_{,\mu }^{A}(x^{\nu },\zeta )={\cal Y}_{,\mu }^{A}+\zeta \bar{%
{\cal N}}_{\,\,\,,\mu }^{A}(x^{\nu }),  \label{1-3}
\end{equation}%
In a similar manner, one can find that since the vectors
$\bar{{\cal N}}^{A}$ depend only on the local coordinates $x^{\mu
}$, they do not propagate along extra dimension
\begin{equation}
{\cal N}^{A}(x^{\mu })=\bar{{\cal N}}^{A}+\zeta [\bar{{\cal
N}},\bar{{\cal N}}]^{A}=\bar{{\cal N}}^{A}. \label{1-4}
\end{equation}
The above assumptions lead to the embedding equations of the
perturbed geometry
\begin{equation}
{\cal G}_{\mu \nu }={\cal G}_{AB}{\cal Z}_{\,\,\ ,\mu }^{A}{\cal
Z}_{\,\,\ ,\nu }^{B},\hspace{0.5cm}{\cal G}_{\mu 4}={\cal
G}_{AB}{\cal Z}_{\,\,\ ,\mu
}^{A}{\cal N}^{B},\hspace{0.5cm}{\cal G}_{AB}{\cal N}^{A}%
{\cal N}^{B}={\cal G}_{44}.  \label{1-5}
\end{equation}%
If we set ${\cal N}^{A}=\delta _{4}^{A}$, the metric of the bulk
space can be written in the following matrix form (Gaussian frame)
\begin{equation}
{\cal G}_{AB}=\left( \!\!\!%
\begin{array}{cc}
g_{\mu \nu }&0 \\
0 & 1
\end{array}%
\!\!\!\right) ,  \label{1-6}
\end{equation}
where $g_{\mu \nu }$ is the metric of the perturbed brane
\begin{equation}
g_{\mu \nu }=\bar{g}_{\mu \nu }-2 \zeta \bar{K}_{\mu \nu }+\zeta^2
\bar{g}^{\alpha \beta }\bar{K}_{\mu \alpha}\bar{K}_{\nu \beta},
\label{1-7}
\end{equation}%
and
\begin{equation}
\bar{K}_{\mu \nu}=-{\cal G}_{AB}{\cal Y}_{\,\,\,,\mu }^{A}{\cal
N}_{\,\,\ ;\nu }^{B},  \label{1-8}
\end{equation}
represents the extrinsic curvature of the original brane (second
fundamental form). Any fixed $\zeta$ signifies a new perturbed
geometry, enabling us to define an extrinsic curvature similar to
the original one by
\begin{equation}
{K}_{\mu \nu}=-{\cal G}_{AB}{\cal Z}_{\,\,\ ,\mu }^{A}{\cal N}%
_{\,\,\ ;\nu }^{B}=\bar{K}_{\mu \nu }-\zeta\bar{K}_{\mu \gamma }%
\bar{K}_{\,\,\ \nu }^{\gamma }. \label{1-9}
\end{equation}
Equations (\ref{1-7}) and (\ref{1-9}) are important in obtaining
the Einstein equation on the perturbed brane in our approach and
show the relation of original and perturbed branes. In the Induced
Matter approach, the Einstein equation in the bulk is written in
the form
\begin{equation}
{\cal R}_{AB}=0,  \label{1-10}
\end{equation}%
where ${\cal R}_{AB}$ is the Ricci tensor of the $5D$ bulk space.
To obtain the effective field equations in $4D$, let us start by
contracting the Gauss-Codazzi equations \cite{14a}
\footnote{Eisenhart's  convention \cite{14a} has been used in
defining the Riemann tensor.}
\begin{equation}
R_{\alpha \beta \gamma \delta }=2\epsilon K_{\gamma [\alpha
}K_{\beta ]\delta
}+{\cal R}_{ABCD}{\cal Z}_{,\alpha }^{A}{\cal Z}_{,\beta }^{B}{\cal Z}%
_{,\gamma }^{C}{\cal Z}_{,\delta }^{D}\label{1-11}
\end{equation}%
and
\begin{equation}
2K_{\mu \lbrack \nu ;\rho ]}={\cal R} _{ABCD}{\cal Z}_{,\mu
}^{A}{\cal N}^{B}{\cal Z}_{,\nu }^{C}{\cal Z} _{,\rho }^{D}.
\label{1-12}
\end{equation}
where ${\cal R}_{ABCD}$ and $R_{\alpha \beta \gamma \delta }$ are
the Riemann curvature of the bulk and perturbed brane
respectively. Now by contracting Gauss equations (\ref{1-11}),
decomposing Riemann tensor of the bulk space into the Weyl and
Ricci tensors and Ricci scalar and using equation (\ref{1-10}),
the Einstein equation induced on the perturbed brane becomes
\begin{equation}
G_{\mu \nu }=Q_{\mu \nu }-{\cal E}_{\mu \nu },  \label{1-13}
\end{equation}%
where ${\cal E}_{\mu \nu }={\cal C}_{ABCD}{\cal Z} _{,\mu
}^{A}{\cal N}^{B}{\cal N}^{C}{\cal Z}_{,\nu }^{D}$ is the electric
part of the Weyl tensor ${\cal C}_{ABCD}$ of bulk space. One can
directly show that $Q_{\mu\nu}$ is independently a conserved
quantity, that is $Q_{\,\,\,\,\,\,;\mu }^{\mu \nu }=0$. All of the
above quantities in equation (\ref{1-13}) are obtained on the
perturbed brane, since in the spirit of IMT the matter field
cannot exactly be confined to the original non perturbed brane.
Hence from a $4D$ point of view, the empty $5D$ equations look
like the Einstein equation with induced matter. The electric part
of the Weyl tensor is well known from the brane point of view. It
describes a traceless matter, denoted by dark radiation or Weyl
matter. Since $Q_{\mu \nu }$ is a conserved quantity, according to
the spirit of IMT \cite{9a}, it should be related to the ordinary
matter as partly having a geometrical origin
\begin{equation}
Q_{\mu\nu} = -8\pi G T_{\mu\nu}.\label{1-14}
\end{equation}
 According to Wesson's results, the IMT
can contain quantum effects \cite{10a,11a}. Now, the question arises
as to whether  the classical fluctuations of the brane can relate to
the quantum fluctuations of the matter field. In other words, can
induced field equation (\ref{1-13}) describe the semiclassical
Einstein equation? In the following we focus on deriving the
semiclassical corrections to the Einstein equation through the IMT.

 One knows that quantum corrections to
general relativity are expected to be important in regimes where
the curvature is near the Planck scale ($l_{pl}=1.6\times 10^{-33}
cm$). In a regime where the curvature approaches but always
remains less than the Planck length, a semiclassical approximation
to the full theory of quantum gravity should be sufficient.  It
may seem that if the brane perturbations are of the order of the
Planck length, equation (\ref{1-13}) may relate to the
semiclassical description of quantum gravity. As we know the
effective size of the extra dimension which should be smaller than
$0.2$  $mm$ can be obtained from
$$L=\frac{M_p^2}{M^3_*},$$ provided $M_*>2\times10^8GeV$ \cite{15a}.
 Here $M_p$ and $M_*$ are the
Planck mass and the fundamental scale of the energy in the bulk
space respectively. On the other hand, the standard model fields are
usually confined to the brane within some localization width i.e,
the brane width \cite{16a,17a}. Similarly, in Induced Matter theory,
if the induced matter satisfies the restricted energy condition, the
particles will be stabilized around the original brane \cite{18a}.
The size of the fluctuations of the induced matter corresponds with
the width of the brane. Since within the ordinary scales of energy
we do not see the disappearance of particles, one may assume the
fluctuations of the matter field exist only around the original
brane. In other words, if the brane width is $d$, it means that
brane localized particles probe this length scale across the brane
and therefor the observer cannot measure the distance on the brane
to a better accuracy than $d$. In braneworld models with large extra
dimensions, usually the width of the brane should be in order of or
less than $TeV^{-1}$. On the other hand, the observational data
constrains the brane width to be in order of planck length, see
\cite{19a} and references therein. Hence, in this paper according to
\cite{20a} we assume that the size of the fluctuations of the brane
(the width of the brane) is in order of Planck length which is much
smaller than the effective size of the extra dimension $L$. This
assumption may help us to investigate the correct quantum
phenomenology in IMT. In the following section we proceed to study
the Einstein equation on the perturbed bran applying a simple
gravitational model and derive the semiclassical corrections to the
Einstein equation.

\section{FRW model with quantum corrections}

Consider a FRW universe embedded (as a non perturbed brane) in an
5D flat bulk space so that  the extra dimension is spacelike. The
FRW line element is written as
\begin{equation}
ds^{2}=-dt^{2}+a^{2}(t)\left[ \frac{dr^{2}}{1-kr^2}+r^{2}(d\theta
^{2}+\sin ^{2}\theta d\varphi ^{2})\right] , \label{1-15}
\end{equation}%
where $k $ takes the values $\pm1$ or $0$ and $a(t)$ is the scale
factor. Now, we proceed to analyze the Einstein equation
(\ref{1-13}) on the perturbed brane. To do this, we first compute
the extrinsic curvature through solving the Codazzi equations
(\ref{1-12}) \cite{21a}
\begin{eqnarray}
\bar{K}_{00}=-\frac{1}{\dot{a}}\frac{d}{dt}\left(
\frac{b}{a}\right),\hspace{2cm} \nonumber\\
\bar{K}_{\alpha\beta}=\frac{b}{a^{2}}g_{\alpha\beta}, \hspace{1.3
cm} \alpha,\beta \neq 0.\label{1-16}
\end{eqnarray}%
Here, $b$ is an arbitrary function of $t$. Consequently, the
components of $\bar {Q}_{\mu \nu }$ become\\

\begin{eqnarray}\label{1-17}
\bar{Q}_{00}=-\frac{3}{a^{4}}b^2,\hspace{4cm}\nonumber\\
\bar{Q}_{\alpha\beta}=\frac{1}{a^4}\left( 2\frac{b\dot{b}}{H}-
b^2\right) g_{\alpha\beta} \hspace{.5 cm}\alpha,\beta\neq 0,
\end{eqnarray}
where $H=\frac{\dot{a}}{a}$ is the Hubble parameter. Note that the
electric part of the Weyl tensor vanishes in the case of the flat
bulk. To proceed with geometrical interpretation of the energy
momentum tensor, let us consider an analogy between $\bar{Q}_{\mu
\nu }$ and a simple example of matter consisting of free radiation
field and cosmological constant $\Lambda$ , that is
\begin{equation}\label{1-18}
\bar{Q}_{\mu \nu }=-8\pi G T_{\mu \nu }+\Lambda
g_{\mu\nu},\hspace{2.3cm}
\end{equation}
where
\begin{equation}\label{1-19}
  T_{\mu \nu }=(\rho +p)u_{\mu}u_{\nu }+p g_{\mu\nu}, \hspace{1cm}
  p=\frac{1}{3}\rho.
\end{equation}
 Using equations (18-20) the energy density and cosmological constant take the
following forms
\begin{eqnarray}\label{1-20}
\rho =\frac{3}{16\pi G a^4}(-\frac{b\dot{b}}{H}+2b^2)\nonumber,\\
 \Lambda=\frac{3}{2}\frac{b\dot{b}}{a^4H}.\hspace{2.3cm}
\end{eqnarray}%
For a radiative universe, which we have $\rho a^4= \rho_0 a_0^4$,
equations (\ref{1-20}) lead to
\begin{equation}\label{1-21}
b=\sqrt{\frac{\rho^*}{2}+\frac{\Lambda a^4}{3}},
\end{equation}
where $\rho_0$ and $a_0$ are the radiation density and the scale
factor at the present epoch and
\begin{equation}\label{1-21a}
\rho^*=\frac{16\pi G}{3}\rho_0 a_0^4.
\end{equation}
 Using equations (19-22), the $4D$ induced components of the
Einstein equation (\ref{1-13}) on the non perturbed brane become
\begin{eqnarray}
3\frac{\dot{a}^2+k}{a^2}=\frac{2a^4\Lambda+3\rho^*}{2a^4},\label{1-22}
\end{eqnarray}
\begin{eqnarray}
\frac{2a\ddot{a}+k+\dot{a}^2}{1-kr^2}=\frac{2a^4\Lambda-\rho^*}{2a^2(1-kr^2)}.\label{1-23}
\end{eqnarray}
Now, one can obtain the components of the tensors $G_{\mu\nu}$ and
$Q_{\mu\nu}$ on the perturbed brane using the components of the
metric and the extrinsic curvature of the perturbed brane through
(\ref{1-7}) and (\ref{1-9}). After some manipulations we obtain
\begin{eqnarray}
Q_{00}=-\frac{2a^4\Lambda+3\rho^*}{2a^4}-\frac{\rho^*\sqrt{12a^4\Lambda+18\rho^*}}{a^6}\zeta\hspace{4cm}\\\nonumber
-\frac{243\rho^{*4}+378a^4\Lambda\rho^{*3}+360a^8\Lambda^2\rho^{*2}+72a^{12}\Lambda^3\rho^*+16a^{16}\Lambda^4}
{3a^8(2a^4\Lambda+3\rho^{*})^{2}}\zeta^2,\vspace{2cm}\label{1-24}
\end{eqnarray}
\begin{eqnarray}\vspace{4cm}
  Q_{\alpha\beta}=[\frac{2a^4\Lambda-\rho^*}{2a^4}+\frac{2(3\rho^*-2a^4\Lambda)}
{a^6\sqrt{12a^4\Lambda+18\rho^*}}\zeta\hspace{7cm}\\\nonumber
  +\frac{54\rho^{*4}+126a^4\Lambda\rho{^*3}
+228a^8\Lambda^2\rho^{*2}
+40a^{12}\Lambda^3\rho^*+16a^{16}\Lambda^4}{3a^8(2a^4\Lambda+3\rho^*)^2}\zeta^2
]g_{\alpha\beta} \hspace{.5 cm}\alpha,\beta\neq 0
 \label{1-25},
\end{eqnarray}
and
\begin{equation}
\hspace{1cm}G_{00}=-3\frac{\dot{a}^2+k}{a^2}-36\frac{(\dot{a}^2+k)\rho^*}{a^4\sqrt{12a^4\Lambda+18\rho^*}}\zeta
-12\rho^*\frac{3\rho^*\ddot{a}^2+\ddot{a}^2a^4\Lambda+3\rho^*
k+k\Lambda a^4}{(2a^4\Lambda+3\rho^*)a^6}\zeta^2,\label{1-26}
\end{equation}
\begin{eqnarray}
G_{\alpha\beta}=\left[\frac{2a\ddot{a}+k+\dot{a}^2}{a^2}-\frac{12\rho^*\ddot{a}}{a^3\sqrt{12a^4\Lambda+18\rho^*}}
\zeta+\frac{2\rho^*\ddot{a}(3\rho^*-2a^4\Lambda)}{(2a^4\Lambda+3\rho^*)a^5}\zeta^2\right]g_{\alpha\beta}\hspace{.5
cm}\alpha,\beta\neq 0.\label{1-27}
\end{eqnarray}
Using the above equations we can derive the $(0,0)$ component of
the Einstein equation on the perturbed brane up to third order in
$\zeta$
\begin{equation}
-3\frac{\dot{a}^2+k}{a^2}=-\Lambda-\frac{3\rho^*}{2a^4}-\frac{(2a^4\Lambda+3\rho^*)^2}{3a^8}\zeta^2
+\frac{4a^4\Lambda\rho^*}{a^8}\zeta^2-\frac{144\Lambda^2\rho^{*2}}{3(2a^4\Lambda+3\rho^*)^2}\zeta^2
+ {\cal O}(\zeta^3) .\label{1-28}
\end{equation}
It is obvious that this equation includes some correction terms of
order $\zeta^2$ in comparison with the Einstein equation
(\ref{1-22}) on the original brane. Since the solution of
(\ref{1-28}) is a correction to solution of (\ref{1-22}) in power of
curvature we expect that at late times, when the solution on the
original brane is nearly flat, these corrections will be flat. At
early times, when the curvature is below the Planck scale, we expect
the corrections to be significant and at very early times, when the
original brane curvature is near or above the planck scale, we
expect that this approach will break down because neglected higher
order corrections would dominate.

According to  Wesson \cite{20a}, we assume that the region of the
brane fluctuations is in order of Planck length $(c=1, G=1)$
\begin{equation}
\zeta^2 \sim l^2_{pl} \sim \hbar.\label{1-29}
\end{equation}
Now we can redefine (\ref{1-28}) using (\ref{1-21a}) and
(\ref{1-29})
\begin{eqnarray}\label{1-30a}
   3\frac{\dot{a}^2}{a^2}+\frac{3k}{a^2}-\Lambda-\frac{\kappa\rho_0 a_0^4}{a^4}+
   \frac{\hbar}{3}\frac{8\kappa\Lambda\rho_0a_0^4}{a^4}\hspace{4cm}\vspace{2cm}\\\nonumber -
   4\hbar\left[\frac{\Lambda^2}{3}+\frac{2\kappa\Lambda\rho_0
   a_0^4}{3a^4}+
   \frac{\kappa^2\rho_0^2 a_0^8}{3a^8}\right]-\frac{16}{3}\hbar\left[\frac{\kappa\Lambda\rho_0 a_0^4}
   {(a^4\Lambda+\kappa\rho_0 a_0^4)}\right]^2
   =0,
\end{eqnarray}
 where $\kappa=8\pi$. One may compare equation (\ref{1-30a}) with the same result which
  is obtained through the semiclassical approach to Einstein
equations in \cite{22a}. There, the semiclassical Einstein
equation for the conformally flat spacetimes take the form
\begin{equation}\label{1-32}
  G_{\mu\nu}+\Lambda g_{\mu\nu}+\alpha_1 \hbar\hspace{0.4cm}
  H^{\hspace{-0.7cm}(1)}_{\mu\nu}+\alpha_3\hbar\hspace{0.4cm}
  H^{\hspace{-0.7cm}(3)}_{\mu\nu}=-\kappa T_{\mu\nu}.
\end{equation}
In this theory the parameters $\alpha_1$ and $\alpha_3$ depend on
the particular form of matter and regularization scheme and
\begin{eqnarray}\label{1-33}
    H ^{\hspace{-0.7cm}(1)}_{\mu\nu}=\frac{1}{2}R^2g_{\mu\nu}-2RR_{\mu\nu}-2\Box R
    g_{\mu\nu}+2\nabla_\mu\nabla_\nu R,\hspace{-3cm}
\end{eqnarray}

\begin{eqnarray}
H^{\hspace{-0.7cm}(3)}_{\mu\nu}=-\frac{1}{2}R^2g_{\mu\nu}+R^{\rho\sigma}R_{\rho\mu\sigma\nu}.\label{1-34}
\end{eqnarray}
 In this approach, although the quantum corrections contain up
to fourth order derivatives of metric, the physically relevant
equations are obtained using the self-consistent method
\cite{22a}. In a FRW background metric, the $(0,0)$ component of
the Einstein equations with first-order semiclassical quantum
corrections is obtained in the form \cite{22a}
\begin{eqnarray}\label{1-35}
   3\frac{\dot{a}^2}{a^2}+\frac{3k}{a^2}-\Lambda-\frac{\kappa\rho_0
   a_0^4}{a^4}-\alpha_1
  \hbar\frac{8\kappa\Lambda\rho_0a_0^4}{a^4}+\alpha_3\hbar
  \left[\frac{\Lambda^2}{3}+\frac{2\kappa\Lambda\rho_0
   a_0^4}{3a^4}+
   \frac{\kappa^2\rho_0^2 a_0^8}{3a^8}\right]=0.
\end{eqnarray}
Comparing (\ref{1-30a}) and (\ref{1-35}) indicates that if we choose
the parameter $\alpha_1=-\frac{1}{3}$ and $\alpha_3=-4$ for the
radiation matter, the result of the both methods are the same up to
the last term in (\ref{1-30a}). This term cannot be obtained using
the self-consistent method. This comparison may imply that one can
interpret (\ref{1-30a}) as the Einstein equation with the first
order quantum corrections.

In $4D$, for conformally flat classical background, when the quantum
state is constructed from the conformal vacuum (conformally trivial
case), the semiclassical corrections to  Einstein equation are
completely determined by local geometry and there is no additional
non-local contribution to the stress tensor \cite{3a}. Computing the
semiclassical Einstein equation with quantum corrections in a
general case is difficult because it includes the state dependent
part of the expectation value of quantum matter fields. As was
indicated we can derive the local corrections to the Einstein
equations via IMT with a 5D flat bulk space (${\cal E}_{\mu\nu}=0$).
Studying a general case in Induced Matter approach may be possible
via computing the Weyl tensor ${\cal E}_{\mu\nu}$, that carriers
non-local effects onto the brane \cite{23a}. Hence, one can say the
non-local quantum contributions to the stress tensor from the point
of view of IMT may be related to the non-local effects of bulk
space. One simple case is when we attempt to obtain the quantum
effects in black holes. In this case the quantum effects diverge
near the singularity. This allows for the possibility that black
holes without singularities might occur in nature \cite{23a}. On the
other hand, it is possible to obtain non-singular black hole
solutions in the brane world model by solving the effective field
equations for the induced metric on the brane \cite{24a}. Hence
obtaining the quantum corrections via IMT in the case of non
vanishing ${\cal E}_{\mu\nu}$ is an interesting subject worth
studying.

\section{Conclusions}
We have considered a FRW universe embedded in a 5D flat bulk space
with a space like extra dimension. We have shown that the
corrections to Einstein equation through the fluctuations of brane
can  correspond to the semiclassical quantum corrections to the
Einstein equation for the conformally trivial case. Although this
correspondence was obtained in a conformally flat background, it may
be generalized to a general case by considering the non-local
effects of ${\cal E}_{\mu\nu}$.

\section*{acknowledgements}
The authors thank prof. H. R. Sepangi for reading the manuscript.
This work has been supported by Research Institute for Astronomy and
Astrophysics of Maragha.

\end{document}